\documentstyle[12pt]{article}
\renewcommand{\theequation}{\arabic{equation}}
\renewcommand{\thesection}{\arabic{section}}
\renewcommand{\thefootnote}{\fnsymbol{footnote}}
\newcommand{\bea}{\begin{eqnarray}}
\newcommand{\ena}{\end{eqnarray}}
\newcommand{\vs}[1]{\vspace{#1 mm}}

\newcommand{\z}{\omega}

\newcommand{\PL}[1]{Phys.\ Lett.\ {\bf #1}}

\newcommand{\PR}[1]{Phys.\ Rev.\ {\bf #1}}
\newcommand{\PRL}[1]{Phys.\ Rev.\ Lett.\ {\bf #1}}
\newcommand{\PTP}[1]{Prog.\ Theor.\ Phys.\ {\bf #1}}

\newcommand{\EPJ}[1]{Eur.\ Phys.\ J.\ {\bf #1}}

\begin{document}
\noindent
\topmargin 0pt
\oddsidemargin 5mm

\begin{titlepage}
\setcounter{page}{0}
\begin{flushright}
March, 2000\\
OU-HET 343\\
hep-ph/0003139\\
\end{flushright}
\vs{10}
\begin{center}
{\Large{\bf Large CP Violation, Large Mixings of 
Neutrinos and the $Z_3$ Symmetry}}\\
\vs{10}
{\large  
Takahiro Miura\footnote{e-mail address: 
miura@het.phys.sci.osaka-u.ac.jp}, 
Eiichi Takasugi\footnote{e-mail address: 
takasugi@het.phys.sci.osaka-u.ac.jp} 
and Masaki Yoshimura\footnote{e-mail address: 
masaki@het.phys.sci.osaka-u.ac.jp}}
\\
\vs{8}
{\em Department of Physics,
Osaka University \\ Toyonaka, Osaka 560-0043, Japan} \\
\end{center}
\vs{10}
\centerline{{\bf Abstract}}  
We present neutrino mass matrices which predict 
the atmospheric neutrino mixing to be almost maximal, 
$\sin^2 2\theta_{atm}>0.999$, 
as well as the large solar neutrino mixing, 
$8/9>\sin^2 2\theta_{sol}>0.87$, and the large CP violation 
(the CP violation phase in the standard form is maximal 
$\delta=\pi/2$), based on the $Z_3$ symmetry. 

\end{titlepage}

\newpage
\renewcommand{\thefootnote}{\arabic{footnote}}
\setcounter{footnote}{0}

\section{Introduction}

The observation of the atmospheric neutrino by Super-Kamiokande[1] 
has shown the existence of the neutrino masses and the 
neutrino mixings. In particular, the data[1] show that the 
mixing between $\nu_\mu$ and $\nu_\tau$ is favored and 
\bea
\sin^2 2\theta_{atm}\sim 1\;,\;\;
\Delta_{atm}^2 \equiv |m_3^2-m_2^2|\sim 3.5\times 10^{-3}
{\rm eV}^2\;.
\ena
The solar neutrino problem is now considered to be due to 
the $\nu_e$ and $\nu_\mu$ oscillation, but the information 
on masses and mixing angles is ambiguous. Now four 
solutions are possible[2]. The another 
crucial information is given by CHOOZ group[3] that 
gives 
\bea
|(V_{MNS})_{13}|<0.16\;,
\ena
provided that $\Delta_{atm}^2= 3.5\times 10^{-3}
{\rm eV}^2$. 
Here, $V_{MNS}$ is the Maki-Nakagawa-Sakata (MNS)[4] neutrino mixing matrix. 

If we interpret these information in the three generation mixing,  
we find 
\bea
\sin^2 2\theta_{atm}\simeq \sin^2 2\theta_{23} c_{13}^4\;
\ena
and 
\bea
|s_{13}|<0.16\;,
\ena
where $\theta_{ij}$ is the mixing angle between the i-th 
and the j-th mass eigenstate neutrinos, 
$s_{ij}=\sin \theta_{ij}$, $c_{ij}=\cos \theta_{ij}$, 
and we used the standard parameterization of mixing matrix[5] 
given in Appendix A. 

If we obtain the large $\sin^2 2\theta_{atm}$  with 
$|s_{13}|<0.16$, we need $|s_{23}|\sim |c_{23}|\sim 1/\sqrt2$. 
Now we face the following questions: \\
(1) Why $|s_{23}|\sim |c_{23}|\sim 1/\sqrt2$?\\
(2) Why $s_{13}$ is so small? \\
(3) How large is the CP violation phase $\delta$? 

In order to answer these questions, various mixing schemes 
have been proposed. Among them, the bi-maximal mixing 
matrix[6], $(V_{MNS})_{bi}$, and the democratic one[7], 
$(V_{MNS})_{demo}$, predict the large mixing for 
both the solar and the atmospheric neutrino mixings, 
\bea
(V_{MNS})_{bi}=\pmatrix{\frac1{\sqrt{2}}&-\frac1{\sqrt{2}}&0\cr 
            \frac12&\frac12&-\frac1{\sqrt 2}\cr
            \frac12&\frac12&\frac1{\sqrt 2}\cr  }\;,\;
(V_{MNS})_{demo}=\pmatrix{\frac1{\sqrt{2}}&-\frac1{\sqrt{2}}&0\cr 
            \frac1{\sqrt 6}&\frac1{\sqrt 6}&-\frac2{\sqrt 6}\cr
            \frac1{\sqrt 3}&\frac1{\sqrt 3}&\frac1{\sqrt 3}\cr}\;.
\ena
The bi-maximal mixing matrix predicts $\sin^2 2\theta_{sol}=1$ and 
$\sin^2 2\theta_{atm}=1$, and the democratic mixing matrix 
predicts $\sin^2 2\theta_{sol}=1$ and 
$\sin^2 2\theta_{atm}=8/9$. 

In our previous paper[8], we proposed a new type of the neutrino 
mixing matrix based on the democratic-type mass matrix which 
is derived from the $Z_3$ invariant Lagrangian. This model 
predicts the most needed relation $|s_{23}|= |c_{23}|= 1/\sqrt2$ 
and in addition $\delta=\pi/2$, the largest CP violation phase 
in the standard parameterization. 
In this paper, we  examine the 
democratic-type mass matrix further and explore the possibility 
of constructing more predictive models. 

This paper is organized as follows: In Sec.2, we briefly review 
the democratic-type mass matrix and its predictions. We 
explain the origin of the mass matrix based on $Z_3$ symmetry, 
by using the see-saw mechanism.  In Sec.3, we consider the 
one Higgs doublet case and introduce the $Z_3$ symmetry 
breaking terms. The predictions are discussed in detail. 
The summary is given in Sec.4.

\section{The democratic-type mass matrix and the $Z_3$ symmetry}

In this section, we present another view of 
the democratic-type mass matrix and its origin. Throughout of this 
paper, we consider the neutrino mass matrix in the charged 
lepton mass eigenstate basis. 

\noindent
(a) The mixing matrix derived from the 
deformation from the tri-maximal mixing matrix

The tri-maximal mixing matrix was discussed extensively 
by many authors[9] and is defined by 
\bea
(V_{MNS})_{tri}\equiv V_T=\frac{1}{\sqrt{3}}\pmatrix{
1 & 1 & 1 \cr
\z & \z^2 & 1 \cr
\z^2 & \z & 1
}\;,
\ena
where $\omega=e^{i2\pi/3}$ or $\omega=e^{i4\pi/3}$, i.e., 
$\omega^3=1$. 
This model predicts $|(V_{MNS})_{13}|=1/\sqrt{3}$ which conflicts 
with the CHOOZ data[3], but it has an interesting property that 
it predicts the maximal value of the CP violation phase in the 
standard form, $\delta=\pi/2$, and  the maximal CP violation, 
that is, the maximal value of the rephasing invariant Jarlskog
parameter, 
$|J_{CP}|_{tri}=1/6\sqrt{3}$. 

In order to remedy the deficit of the model, 
it may be interesting to consider the deformation from 
the tri-maximal mixing matrix by an orthogonal matrix $O$, 
\bea
V=V_TO\;.
\ena
The orthogonal matrix contains three angles so that naively 
we expect that the CP violation phase is expressed by three angles in a 
complicated expression. Contrarily to this expectation, we found[8]
\bea
|\sin \theta_{23}|=|\cos \theta_{23}|=\frac1{\sqrt 2} \;,
\qquad \delta=\frac{\pi}2\;
\ena
and two Majorana phases are fixed uniquely 
independent of angle parameters in $O$. Explicitly, we found 
that the neutrino mixing matrix 
is
\bea
V_{MNS}=\pmatrix{c_{12}c_{13}&s_{12}c_{13}&-is_{13}\cr
 -\frac{s_{12}-ic_{12}s_{13}}{\sqrt 2}&
 \frac{c_{12}+is_{12}s_{13}}{\sqrt 2}&-\frac{c_{13}}{\sqrt 2}\cr
 -\frac{s_{12}+ic_{12}s_{13}}{\sqrt 2}&
 \frac{c_{12}-is_{12}s_{13}}{\sqrt 2}&\frac{c_{13}}{\sqrt 2}\cr}
 \pmatrix{1&0&0\cr0&1&0\cr0&0&i\cr}\;.
\ena
The diagonal matrix diag$(1,1,i)$ represents the Majorana phase 
matrix which is relevant to the purely lepton number violating 
processes such as the neutrinoless double beta decay[10]. 
One parameter out of three parameters in $O$ is converted to 
the phase  which is  absorbed by the redefinition of charged 
leptons.  The brief derivation is given in Appendix A.  

If we use the CHOOZ constraint, $|(V_{MNS})_{13}|=|s_{13}|<0.16$, 
the model predicts
\bea
\sin^22\theta_{atm}&=&4|(V_{MNS})_{23}|^2(1-|(V_{MNS})_{23}|^2)\nonumber\\
&=& 1-s_{13}^4>0.999  \;,\nonumber\\
\frac{|J_{CP}|_{our\; model}}{|J_{CP}|_{max}} &=&
\frac{3\sqrt 3}2
|\sin 2\theta_{12}s_{13}c_{13}^2|\;,
\ena
where $\sin^2 2\theta_{12}\simeq \sin^2 2\theta_{sol}$ is 
determined by the solar neutrino data. Therefore, in our 
model, the size of the CP violation is determined by 
the solar neutrino mixing angle and $|s_{13}|$  which 
has been bounded by the CHOOZ data. 
If we take the large mixing solution for the solar neutrino 
problem, $\sin^2 2\theta_{sol}=0.9$ and the largest allowed 
value for $s_{13}$, 
$|s_{13}|=0.16$, our model predicts $|J_{CP}|_{our\; model}=0.38
|J_{CP}|_{max}$.

\vskip 2mm
\noindent
(b) The possible origin of the mixing matrix in the form of 
$V=V_TO$
 
Let us consider what kind of mass matrix leads to $V=V_TO$. 
In order to clarify the structure, we change the 
flavor eigenstate basis to the one which is obtained by 
transforming by $V_T$ (hereafter we call it as the $\psi$ basis),
\bea
\pmatrix{\psi_1\cr\psi_2\cr\psi_3\cr}=V_T^\dagger 
\pmatrix{\nu_{e L}\cr \nu_{\mu L}\cr\nu_{\tau L}\cr}
=\frac1{\sqrt{3}}\pmatrix{1&\z^2&\z\cr 1&\z&\z^2\cr 1&1&1\cr}
\pmatrix{\nu_{e L}\cr \nu_{\mu L}\cr\nu_{\tau L}\cr}\;.
\ena
The relation $V=V_TO$  implies that when we look at the mass matrix 
in the $\psi$ basis, the neutrino mass matrix should be a 
real symmetric matrix. Thus, we may parameterize 
\bea
\tilde m_\nu=V_T^T m_\nu V_T= \pmatrix{
   m^0_1+\tilde{m}_1&\tilde m_3&\tilde m_2\cr
   \tilde m_3&m^0_2+\tilde{m}_2&\tilde m_1\cr
   \tilde m_2&\tilde m_1&m^0_3+\tilde{m}_3\cr}\;,
\ena
with the real parameters, $m_i^0$ and $\tilde m_i$. 
Here, $\tilde m_\nu$ is the mass matrix in the $\psi$ basis 
and $m_\nu$ is the one in the flavor eigenstate 
basis. If $\tilde m_\nu$ is a real symmetric matrix, then 
it is diagonalized by the orthogonal matrix $O$ and thus 
the mixing matrix becomes $V_TO$. 

By inverting, we obtain the neutrino mass matrix $m_\nu$ as 
\bea
m_\nu &=&
\frac{m^0_1}3\pmatrix{1&\z^2&\z\cr \z^2&\z&1\cr 
         \z&1&\z^2}+
\frac{m^0_2}3\pmatrix{1&\z&\z^2\cr \z&\z^2&1\cr 
         \z^2&1&\z\cr }+
\frac{m^0_3}3 \pmatrix{1&1&1\cr 1&1&1\cr 1&1&1\cr}
\nonumber\\
&&+
\tilde{m}_1\pmatrix{1&0&0\cr 0&\z&0\cr 0&0&\z^2\cr}+
\tilde{m}_2\pmatrix{1&0&0\cr 0&\z^2&0\cr 0&0&\z\cr}+
\tilde{m}_3\pmatrix{1&0&0\cr 0&1&0\cr 0&0&1\cr}\;,
\ena
which we called the democratic-type mass matrix[8]. 

\vskip 2mm
\noindent
(c) The $Z_3$ symmetric dimension five Lagrangian

We analyze the Lagrangian which gives the democratic-type 
neutrino mass matrix. From the transformation in Eq.(11), 
we define
\bea
\Psi_1&=&\frac1{\sqrt 3}(\ell_e+\z^2 \ell_\mu+\z \ell_\tau)\;,
\nonumber\\
\Psi_2&=&\frac1{\sqrt 3}(\ell_e+\z \ell_\mu+\z^2 \ell_\tau)\;, 
\nonumber\\
\Psi_3&=&\frac1{\sqrt 3}(\ell_e+\ell_\mu+\ell_\tau) \;,
\ena
where $\ell_i$ is the left-handed lepton doublet defined by, 
say,  $\ell_e=(\nu_{eL}, e_L)^T$. With the definition, 
$\Psi_i=(\psi_i, e_i)^T$, the above relation is interpreted 
as the transformation from the flavor eigenstate basis to 
the $\psi$ basis. 

The fields $\Psi_i$ behave as 
irreducible representations of $Z_3$ symmetry under 
the permutation of $\ell_e$, $\ell_\mu$ and $\ell_\tau$,
\bea
\Psi_1\to \z \Psi_1\;,\;\; 
\Psi_2\to \z^2 \Psi_2 \;,\;\;   \Psi_3\to \Psi_3 \;.
\ena
If we introduce two Higgs doublets that behave as 
\bea
H_1 \rightarrow \z^2 H_1 \;,\;  H_2 \rightarrow \z H_2\;, 
\ena
then we can construct the $Z_3$ invariant dimension five 
effective Lagrangian as 
\bea
{\cal L}_{\rm y}
&=&-
  \Biggl(
     (m^0_1+\tilde{m}_1)\overline{(\Psi_1)^C} \Psi_1 
                        \frac{H_1 H_1}{u^2_1}+
     (m^0_2+\tilde{m}_2)\overline{(\Psi_2)^C} \Psi_2 
                        \frac{H_2 H_2}{u^2_2}  \nonumber\\
     &&\mbox{ }
      +(m^0_3+\tilde{m}_3)\overline{(\Psi_3)^C} \Psi_3 
                        \frac{H_1 H_2}{u_1u_2} \Biggr)\nonumber\\
&&-2\left(
     \tilde{m}_1\overline{(\Psi_2)^C} \Psi_3 \frac{H_1 H_1}{u^2_1}+
     \tilde{m}_2\overline{(\Psi_1)^C} \Psi_3 \frac{H_2 H_2}{u^2_2}+
     \tilde{m}_3\overline{(\Psi_1)^C} \Psi_2 \frac{H_1 H_2}{u_1u_2}  
  \right)\;,
\ena
where $u_i$ is the vacuum expectation value of the neutral 
component of $H_i$. After the Higgs fields acquire 
the vacuum expectation values, the neutrino 
mass matrix in the $\psi$ basis defined by Eq.(12) is obtained.

\vskip 2mm
\noindent
(d) The $Z_3$ symmetric Lagrangian and the see-saw 
mechanism

In addition to the $\Psi_i$ and two Higgs doublets, $H_1$ 
and $H_2$, we introduce the right-handed 
neutrinos, $\nu_{eR}$, $\nu_{\mu R}$ and $\nu_{\tau R}$ 
and one more Higgs $H_3$ which behave under the $Z_3$ as 
\bea
\nu_{eR} \to \nu_{eR}\;,\;
\nu_{\mu R}\to \z\nu_{\mu R}\;,\; 
\nu_{\tau R}\to \z^2 \nu_{\tau R}\;,
\ena
\bea
H_3 \rightarrow  H_3\;. 
\ena

Now the $Z_3$ invariant Yukawa interaction and 
the Majorana mass term for the right-handed neutrinos are given by   
\bea 
{\cal L}_{ D}
&=& -(a\overline{\nu_e}_{ R}\frac{H_1}{u_1}+
b\overline{\nu_{\tau}}_{ R}\frac{H_2}{u_2} +d' 
\overline{\nu_{\mu}}_{ R}\frac{H_3}{u_3} )\Psi_1\nonumber\\
&&-(c\overline{\nu_{\mu}}_{ R}\frac{H_1}{u_1}+
d\overline{\nu_e}_{ R}\frac{H_2}{u_2}+
f'\overline{\nu_\tau}_{ R}\frac{H_3}{u_3} )\Psi_2 \nonumber\\
&&-(e\overline{\nu_{\tau}}_{ R}\frac{H_1}{u_1} +
f\overline{\nu_{\mu}}_{ R} \frac{H_2}{u_2}+
b'\overline{\nu_{e}}_{ R}\frac{H_3}{u_3})\Psi_3 +{\rm h.c.}\;,
\ena
where $u_i$ is the vacuum expectation value of $H_i$ and 
\bea
{\cal L}_{ R}
=-M \overline{(\nu_{e R})^C}\nu_{e R}
 - M'\left( \overline{(\nu_{\mu R})^C}\nu_{\tau R}
 +\overline{(\nu_{\tau R})^C}\nu_{\mu R}\right)\;.
\ena

After the Higgs fields acquire the vacuum expectation values, 
the Dirac mass term, 
$m_D$ in the $\psi$ basis and Majorana mass term 
$M_R$ for $\nu_R$ are given by  
\bea
m_D=\pmatrix{a&d&b'\cr d'&c&f\cr b&f'&e\cr}\;,\;\;
M_R=\pmatrix{M&0&0\cr 0&0&M'\cr 0&M'&0}\;.
\ena
By the see-saw mechanism, the neutrino mass matrix for 
the left-handed neutrinos in the $\psi$ basis is given by 
\bea
\tilde m_\nu=-m_D^TM_R^{-1}m_D=-
\pmatrix{\frac{a^2}{M}+\frac{2bd'}{M'}
  &\frac{ad}{M}+\frac{bc}{M'}+\frac{d'f'}{M'}&
\frac{ab'}{M}+\frac{bf}{M'}+\frac{d'e}{M'}\cr
 \frac{ad}{M}+\frac{bc}{M'}+\frac{d'f'}{M'}
& \frac{d^2}{M}+\frac{2cf'}{M'}
& \frac{b'd}{M}+\frac{ce}{M'}+\frac{ff'}{M'}\cr
\frac{ab'}{M}+\frac{bf}{M'}+\frac{d'e}{M'}
& \frac{b'd}{M}+\frac{ce}{M'}+\frac{ff'}{M'}
&\frac{b'^2}{M}+\frac{2fe}{M'}\cr}
\ena
If we parameterize
\bea
&&m_1^0+\tilde m_1=- \left(\frac{a^2}{M}+2\frac{bd'}{M'}\right)\;\;
\qquad  \tilde m_1=- \left(\frac{b'd}{M}+\frac{ce}{M'}+\frac{ff'}{M}\right)
\nonumber\\
&&m_2^0+\tilde m_2=- \left(\frac{d^2}{M}+2\frac{cf'}{M'}\right)\;\;
\qquad  \tilde m_2=- \left(\frac{ab'}{M}+\frac{bf}{M'}+\frac{d'e}{M'}\right)
\nonumber\\
&&m_3^0+\tilde m_3=- \left(\frac{b'^2}{M}+2\frac{fe}{M'}\right)\;\;
\qquad  \tilde m_3=- \left(\frac{ad}{M}+\frac{bc}{M'}+\frac{d'f'}{M'}\right)\;,
\ena
we obtain the neutrino mass matrix $\tilde m_\nu$ in the 
$\psi$ basis 
given in Eq.(12). The ansatz is that all 
parameters $a$, $b$, $c$, $d$, $e$, $f$, $b'$, 
$d'$, $f'$, $M$ and $M'$ are real. 
The Lagrangian in Eqs.(20) and (21) are the most general one 
to derive the democratic mass matrix, although it contains 
redundant parameters. 

The minimal model is those which contain only two Higgs doublets, 
say, $H_1$ and $H_2$, where $\tilde m_\nu$ contains six independent 
parameters to specify $m_i^0$ and $\tilde m_i$.

\section{A restricted model -One Higgs case-}
 
In order to construct the more predictive mass matrix, 
we try to reduce the number of parameters in the democratic-type 
mass matrix, $m_i^0$ and $\tilde m_i$. 
Here, we consider the possibility of one Higgs model. 

If we keep only $H_1=H$ with $\langle H \rangle =u$ in the Lagrangian
(20), 
then  
\bea 
{\cal L}_{D}
= -\frac1{u}\left( a\overline{\nu_e}_{R} H\Psi_1
+c\overline{\nu_{\mu}}_{ R} H\Psi_2 
+e\overline{\nu_{\tau}}_{ R} H\Psi_3\right) 
+{\rm h.c.}\;.
\ena
Then,  the Dirac mass term in the $\psi$ basis is 
$m_D={\rm diag}(a,c,e)$.  After the see-saw mechanism, 
the left-handed neutrino mass matrix in the $\psi$  basis is 
\bea 
\tilde{m}_{\nu}= -m_D^T M_R^{-1}m_D=
\pmatrix{
   m^0_1+\tilde{m}_1&0&0\cr
   0&0&\tilde m_1\cr
   0&\tilde m_1&0\cr}\;.
\ena
In the flavor eigenstate basis 
\bea
m_\nu =V_T^*\tilde m_\nu V_T^\dagger=
\frac{m^0_1}3\pmatrix{1&\z^2&\z\cr \z^2&\z&1\cr 
         \z&1&\z^2}+
\tilde{m}_1\pmatrix{1&0&0\cr 0&\z&0\cr 0&0&\z^2\cr}\;.
\ena
 
From Eq.(26), we see that the neutrino mass matrix $m_\nu$ is 
diagonalized by 
\bea
V_1=V_T\pmatrix{1&0&0\cr0&\frac1{\sqrt 2}&-\frac1{\sqrt 2}\cr
           0&\frac1{\sqrt 2} & \frac1{\sqrt 2}\cr}\;.
\ena  
This mixing scheme is interesting, although this model is 
unrealistic because it predicts the degenerate mass, 
$m_2=-m_3$. 

\vskip 2mm
\noindent
(a) The model with the $Z_3$ symmetry breaking terms

We introduce the $Z_3$ symmetry breaking terms by 
assuming that they respect the $Z_2$ symmetry which is 
defined by 
\bea
\Psi_1 \to -\Psi_1 \;,\;\; \nu_{e R} \to -\nu_{e R}\;, 
\ena 
and  all other fields are unchanged. 
Then, the $Z_2$ invariant Lagrangian is given by
\bea 
{\cal L}_{SB}
&=& -\frac1{u}(a\overline{\nu_e}_{\rm R}H\Psi_1
+c\overline{\nu_{\mu}}_{\rm R}H\Psi_2
+e\overline{\nu_{\tau}}_{\rm R}H\Psi_3 )\nonumber\\
&&-\frac1{u}(f\overline{\nu_{\mu}}_{\rm R}H\Psi_3  
+f'\overline{\nu_{\tau}}_{\rm R}H\Psi_2  ) 
\ena
together with the Majorana mass term for the right-handed 
neutrinos given in Eq.(21). This model gives 
the mass matrix in the $\psi$ basis in Eq.(12) with 
non-zero $m_1^0$, $\tilde m_1$, $m_2^0$ and $m_3^0$. 
In other words, in the flavor eigenstate basis, 
it is 
\bea
m_\nu& =&
\frac{m^0_1}3\pmatrix{1&\z^2&\z\cr \z^2&\z&1\cr 
         \z&1&\z^2}+
 \tilde{m}_1\pmatrix{1&0&0\cr 0&\z&0\cr 0&0&\z^2\cr}
 \nonumber\\
&&\quad +\frac{m^0_2}3 \pmatrix{1&\z&\z^2\cr \z&\z^2&1\cr
\z^2&1&\z\cr}
+\frac{m^0_3}3 \pmatrix{1&1&1\cr 1&1&1\cr 1&1&1\cr}\;.
\ena
In the new basis which is obtained from the transformation 
by $V_1$ in Eq.(28), 
the mass matrix is
\bea
V_1^T m_\nu V_1=\pmatrix{m_1^0+\tilde m_1&0&0\cr
   0&\tilde m_1+\frac12(m_3^0+m_2^0)&\frac12(m_3^0-m_2^0)\cr
  0&\frac12(m_3^0-m_2^0)&-\tilde m_1+ \frac12(m_3^0+m_2^0)\cr}
  \;.
\ena
Thus, the mixing matrix which diagonalize $m_\nu$ is given by 
\bea
V= \pmatrix{1&0&0\cr0&\z&0\cr 0&0&\z^2}
 \pmatrix{\sqrt{\frac13}&-\sqrt{\frac23}&0\cr 
            \frac{1}{\sqrt 3}&\frac{1}{\sqrt 6}&-\frac{1}{\sqrt 2}\cr
            \frac{1}{\sqrt 3}&\frac{1}{\sqrt 6}&\frac{1}{\sqrt 2}\cr}
   \pmatrix{1&0&0\cr 0&-1&0\cr 0&0&i\cr}
   \pmatrix{1&0&0\cr 0&c'&-s'\cr 0&s'&c'\cr}\;,
\ena
where $c'=\cos \theta'$ and $s'=\sin \theta'$ and 
\bea
\tan \theta'=\frac{\frac{m_3^0-m_2^0}2}
{\tilde m_1+\sqrt{\tilde m_1^2+\left(\frac{m_3^0-m_2^0}2\right)^2}}\;,
\ena
and neutrino masses are given by 
\bea
m_1&=&m_1^0+\tilde m_1\;,\nonumber\\
m_2&=&\frac{m_3^0+m_2^0}2+\sqrt{\tilde m_1^2+\left(\frac{m_3^0-m_2^0}2
\right)^2}\;,
\nonumber\\
m_3&=&\frac{m_3^0+m_2^0}2-\sqrt{\tilde m_1^2+\left(\frac{m_3^0-m_2^0}2
\right)^2}\;. 
\ena
Here we take the convention, $\tilde m_1>0$, and also 
consider that $\tilde m_1> |(m_3^0+m_2^0)/2|$, 
$|(m_3^0-m_2^0)/2|$, because 
$m_3^0$ and $m_2^0$ are the $Z_3$ symmetry breaking parameters. Then, we
find 
$m_2>0$ and $m_3<0$. 
The parameter $\tilde m_1$ controls the average 
size of neutrino masses, and $m_3^0+m_2^0$ does 
the mass splitting between 
$m_2$ and $m_3$, and the parameter $m_3^0-m_2^0$ 
does the size of $(V_{MNS})_{13}$. 

The MNS mixing matrix is explicitly given by
\bea
V_{MNS}= 
   \pmatrix{\frac1{\sqrt{3}}&-\sqrt{\frac23}c'&i\sqrt{\frac23}s'\cr 
            \frac 1{\sqrt 3}&\frac1{\sqrt 6}(c'+i\sqrt 3 s')
            &-\frac1{\sqrt 6}(\sqrt 3 c'+is')\cr
            \frac1{\sqrt 3}&\frac1{\sqrt 6}(c'-i\sqrt 3 s')
            &\frac1{\sqrt 6}(\sqrt 3 c'-is')\cr}
   \pmatrix{1&0&0\cr 0&-1&0\cr 0&0&i\cr}\;.
\ena
It should be noted that the model predicts $\delta=\frac{\pi}2$ 
and the angle $\theta_{12}$ which is relevant to the solar 
neutrino mixing. We have 
\bea
\sin^2 2\theta_{sol}= \frac 89 c'^2\;,\;\;
\sin^2 2\theta_{atm}= 1-\frac 49 s'^4 \;.
\ena
If we impose the CHOOZ bound, $|\sqrt{2/3}s'|<0.16$, we have 
\bea
\frac89>\sin^2 2\theta_{sol}>0.87\;,\;\;\;\sin^2 2\theta_{atm}>0.999\;.
\ena
Below, we consider the special cases.

\vskip 2mm
\noindent
(b) Some limiting cases

\vskip 2mm
\noindent
(b-1) The model with $m_3^0=m_2^0$

This model is realized by imposing the invariance under the 
permutation in addition to $Z_2$ symmetry as
\bea
\nu_{\mu R} \iff \nu_{\tau R}\;,\;\;\Psi_2 \iff \Psi_3\;.
\ena
Then, the parameters in Eq.(30) are restricted as 
$f=f'$ and $c=e$, and the condition $m_3^0=m_2^0$ is realized. 
From Eq.(34), we have  $\theta'=0$ 
and neutrino masses are
\bea
m_1=m_1^0+\tilde m_1\;,\;\;m_2=m_3^0+\tilde m_1\;,\;\; 
m_3=m_3^0-\tilde m_1\;.
\ena 
Thus, the model predicts the mixing matrix defined in Eq.(28). 
By parameterizing as $V={\rm diag}(1,\z,\z^2)V_{MNS}$, we find
the MNS matrix is 
\bea
V_{MNS}=\pmatrix{\frac1{\sqrt{3}}&-\sqrt{\frac23}&0\cr 
            \frac1{\sqrt 3}&\frac1{\sqrt 6}&-\frac1{\sqrt 2}\cr
            \frac1{\sqrt 3}&\frac1{\sqrt 6}&\frac1{\sqrt 2}\cr}
             \pmatrix{1&0&0\cr 0&-1&0\cr 0&0&i}\;,
\ena
where the matrix diag$(1,-1,i)$ is the Majorana phase matrix. 

This model predicts 
\bea
\sin^2 2\theta_{atm}=1\;,\;\;\sin^2 2\theta_{sol}=\frac89\;.
\ena
This mixing matrix is in contrast to the so called democratic 
mixing matrix[7] which predicts 
$\sin^2 2\theta_{atm}=8/9$ and $\sin^2 2\theta_{sol}=1$.

\vskip 2mm
\noindent
(b-2) The model with $m_2^0=0$ or $m_3^0=0$

One way to realize this model is to consider two Higgs model 
by keeping $H_1$ and $H_2$ in Eq.(20). Then, we impose 
$Z_3\times Z_2$ symmetry. By the $Z_3$ symmetry, we have 
$b'=d'=f'=0$. Then, we impose the $Z_2$ symmetry given in  
Eq.(29) and we find that $b=d=0$. These conditions give 
$m_2^0=0$. Similarly, $m_3^0=0$ case is realized. 
Since both cases give the same mixing matrix so that we consider 
$m_2^0=0$ case. Then, the mixing angle $\theta'$ is completely 
fixed by the ratio of neutrino masses, $m_2>0$ and $m_3<0$. 
Explicitly, we find with $V_{MNS}=V_{SF}{\rm diag}(1,-1,i)$ 
\bea
V_{SF}=\pmatrix{\frac1{\sqrt{3}}&
     - \sqrt{\frac13 \frac{\beta+2}{\beta} }&
     \pm i\sqrt{\frac13\frac{\beta-2}{\beta} } \cr 
            \frac12\sqrt{\frac{\beta+2}{\beta+1} }
            \left[1 \mp i\sqrt{\frac13 \frac{\beta-2}{\beta+2}
}\right]&
            \frac12\sqrt{\frac{\beta}{\beta+1} }
            \left[1\pm i\sqrt{\frac13 \frac{(\beta+2)(\beta-2)}
            {\beta^2} }\right]&
            -\sqrt{\frac13 \frac{\beta+1}{\beta} }\cr
            \frac12 \sqrt{\frac{\beta+2}{\beta+1}}
            \left[1\pm i\sqrt{\frac13 \frac{\beta-2}{\beta+2}}\right]&
            \frac12\sqrt{\frac{\beta}{\beta+1}}
            \left[ 1\mp i\sqrt{\frac13 \frac{(\beta+2)(\beta-2)}
            {\beta^2}}\right]&
            \sqrt{\frac13 \frac{\beta+1}{\beta} }\cr}
\ena
where 
\bea
\beta=\sqrt{\left|\frac{m_2}{m_3}\right|}+\sqrt{\left|\frac{m_3}{m_2}
\right|}\ge 2\;.
\ena
In the limit $m_2=-m_3$, i.e., $\beta= 2$, the mixing matrix 
reduces to the mixing scheme in Eq.(41). Due to the CHOOZ 
bound, $\beta$ must be close to 2. 

This model predicts 
\bea
\sin^22\theta_{sol}&=&\frac49 \frac{\beta+2}{\beta}\;,
  \nonumber\\
\sin^22\theta_{atm}&=&\frac49\frac{(\beta+1)(2\beta-1)}{\beta^2}
\;.
\ena
If we impose the CHOOZ bound
\bea
|(V_{MNS})_{13}|=\sqrt{\frac13\frac{\beta -2}{\beta}}  < 0.16\;,
\ena
$\beta$ is restricted by $2<\beta<2.17$ which means 
\bea
0.85<\sin^22\theta_{sol}<\frac89\;,\;\;
0.999<\sin^22\theta_{atm}<1\;,
\ena
\bea
0.44 < \left|\frac{m_2}{m_3} \right|<2.27\;.
\ena
The effective mass for the neutrinoless double beta decay 
is given by 
\bea
|<m_\nu>|=\frac13 \left| m_1 +\frac{\beta+2}{\beta}m_2
+\frac{\beta-2}{\beta}m_3 \right|
\sim \frac13 |m_1+2m_2|\;.
\ena
If we take the constraint $|<m_\nu>|<0.3$ eV, then we find 
$|<m_\nu>|\sim |m_2|<0.3$ eV for $m_1m_2>0$ and 
$3|<m_\nu>|\sim |m_2|<0.9$ eV for $m_1m_2<0$.

In order to realize the large CP violation, it is required 
that the value of $|(V_{MNS})_{13}|=\sqrt{(\beta-2)/3\beta}$ 
should be as large as its bound by the CHOOZ data. 
This leads to the constraint on the ratio $|m_2/m_3|$. 
Explicitly, we have
\bea
|m_2|=h\sqrt{\frac{\Delta_{atm}^2}{|h^2-1|}}\;,\;\;
|m_3|=\sqrt{\frac{\Delta_{atm}^2}{|h^2-1|}}\;,
\ena
where
\bea
h=\left\{\frac12\left(\beta\pm \sqrt{\beta^2-4}\right) \right\}^2\;.
\ena
If we take $\Delta_{atm}^2=3.5\times 10^{-3}{\rm eV}^2$ 
and the maximal value $|(V_{MNS})_{13}|=0.16$, 
we find 
$(|m_2|,|m_3|)=(6.6, 2.9)\times 10^{-2}{\rm eV}$ or  
${\rm or} (2.9,6.6)\times 10^{-2}{\rm eV}$. 
When these masses are larger than the above values, 
the CP violation becomes smaller.  

The parameter $\tilde m_1$ is used to fix the overall 
normalization of neutrino masses and $m_3^0$ works  
to adjust the ratio $|m_2/m_3|$. The remaining 
parameter $m_1^0$ determines the mass $m_1$, which is 
related to the neutrino squared mass difference for 
the solar neutrino mixing, $\Delta_{sol}^2\equiv 
|m_2^2-m_1^2|$. Since  $\Delta_{sol}^2$ 
is much smaller than $\Delta_{atm}^2$, 
the fine tuning is required to obtain the approximate degeneracy 
between $m_1$ and $m_2$. The parameter 
$m_1^0$  adjusts the splitting between $m_1$ and $m_2$.

\section{Summary}

In our previous paper[8], we proposed the democratic-type 
neutrino mass matrix  and showed that this model predicts 
the quite interesting relations which are crucial to 
give the large atmospheric neutrino mixing and the 
large CP violation which are given in Eq.(9). 

In this paper, we explored the mass matrices which inherit 
the attractive features we found, but give more predictions. 
In order to examine the origin of the democratic-type mass 
matrix further, we considered the Yukawa interaction 
and the Majorana mass matrix for the 
right-handed neutrinos and derive  the 
neutrino mass matrix for the left-handed neutrinos.  

We gave the most general $Z_3$ invariant Lagrangian 
with three Higgs doublets and showed that this Lagrangian 
gives the democratic-type mass matrix. Then, we restricted 
the number of the Higgs doublet to be one. By introducing 
the $Z_3$ symmetry breaking terms but keeping the $Z_2$ symmetry, we found 
the interesting model that predicts the large solar neutrino 
mixing also. The most prominent feature of the present model 
is the prediction that $\sin^2 2\theta_{atm}$ is very close to 
unity and also the large CP violation (the CP violation 
phase in the standard form $\delta =\pi/2$). These two features 
will be tested as unambiguous signals of the model.

We expect that our mass matrix is obtained at the right-handed 
mass $M_R$ scale. We did not discuss the effect of the 
renormalization in this paper. If $m_1m_2<0$, our predictions 
are stable, while the substantial dependence is expected 
for $\sin^2 2\theta_{sol}$ for $m_1m_2>0$ because of the 
degeneracy. The possibility that the large 
$\sin^2 2\theta_{sol}$ at $M_R$ reduces to the small value 
which explains the small mixing angle MSW solution is under 
estimation and will be published elsewhere.

\vskip 5mm
{\Huge Acknowledgment} 
This work is supported in part by 
the Japanese Grant-in-Aid for Scientific Research of
Ministry of Education, Science, Sports and Culture, 
No.11127209.
 
\newpage

%Appendix format setting
\setcounter{section}{0}
\renewcommand{\thesection}{\Alph{section}}
\renewcommand{\theequation}{\thesection .\arabic{equation}}
\newcommand{\apsc}[1]{\stepcounter{section}\noindent
\setcounter{equation}{0}{\Large{\bf{Appendix\,\thesection:\,{#1}}}}}

\apsc{Derivation of Eq.(8)}

Let us consider
\bea
V&\equiv& V_T O\nonumber\\
&=&\frac1{\sqrt 3} 
    \pmatrix{O_{11}+O_{21}+O_{31}& O_{12}+O_{22}+O_{32}&
    O_{13}+O_{23}+O_{33}\cr
    \z O_{11}+\z^2 O_{21}+O_{31}& \z O_{12}+\z^2 O_{22}+O_{32}& 
    \z O_{13}+\z^2 O_{23}+O_{33}\cr
     \z^2 O_{11}+\z O_{21}+O_{31} & \z^2 O_{12}+\z O_{22}+O_{32} &
      \z^2 O_{13}+\z O_{23}+O_{33} \cr}\;,
      \nonumber\\
\ena
where $V_T$ is the tri-maximal mixing matrix in Eq.(6) and 
$O$ is the orthogonal matrix. Then, we find the constraints
\bea
V_{2j}=V_{3j}^*\;,\qquad (j=1,2,3)
\ena
because $O_{ij}$ are real parameters. Now we define the 
standard form of the mixing matrix as 
\bea
V_{SF}= 
 \pmatrix{c_{12}c_{13}&s_{12}c_{13}&s_{13}e^{-i\delta}\cr
 -s_{12}c_{23}-c_{12}s_{23}s_{13}e^{i\delta}&
  c_{12}c_{23}-s_{12}s_{23}s_{13}e^{i\delta}& s_{23}c_{13}
  \cr
  s_{12}s_{23}-c_{12}c_{23}s_{13}e^{i\delta}&
  -c_{12}s_{23}-s_{12}c_{23}s_{13}e^{i\delta}& c_{23}c_{13}
  \cr}\;,
\ena
where $s_{ij}=\sin \theta_{ij}$ and $c_{ij}=\cos \theta_{ij}$. 
Since 
$V$ is related to $V_{SF}$ by the phase transformation as 
$V=PV_{SF}P'$, the constraints are given by 
$|(V_{SF})_{2j}|=|(V_{SF})_{3j}|$. From the constraint with 
$j=3$, we obtain $|s_{23}|= |c_{23}|$. From $j=1$ and $j=2$, 
together with $|s_{23}|= |c_{23}|$, we find $\cos \delta=0$.


\newpage
\begin{thebibliography}{99}
\bibitem{SuperK-Atm-1999-2}
A.~Mann, 
talk given at {\it 19th International Symposium on Lepton and Photon 
Interactions at High Energies (LP 99), Stanford, CA, 9-14 Aug 1999};\\
hep-ex/9912007.

%%% Super-K
\bibitem{SuperK-Solar-1999-2}
Y.~Suzuki, 
talk given at {\it 19th International Symposium on Lepton and Photon 
Interactions at High Energies (LP 99), Stanford, CA, 9-14 Aug 1999}.

%%% CHOOZ
\bibitem{CHOOZ-1999} 
M.~Apollonio {\it et al.}, \PL{B466} (1999) 415.

%%% MNS %%%
\bibitem{MNS}
Z.~Maki, M.~Nakagawa and S.~Sakata, 
\PTP{28} (1962) 870.

%%% PDG %%%
\bibitem{PDG-1998} C.~Caso {\it et al.}, \EPJ{C3} (1998) 1.

%%% Bi-Maximal %%%
\bibitem{BiMax} 
F.~Vissani, hep-ph/9708483;\\
V.~Barger, S.~Pakvasa, T.J.~Weiler, and K.~Whisnant,
\PL{B437} (1998) 107;\\
A.J.~Baltz, A.S.~Goldhaber, and M.~Goldhaber, \PRL{81} (1998) 5730.

%%% Democratic-Mixing %%%
\bibitem{DemoMix} 
H.~Harari, H.~Kaut and J.~Weyers, \PL{B78} (1978) 459;\\
Y.~Koide, \PRL{D39} (1989) 1391;\\
H.~Fritzsch and Z.Z.~Xing, \PL{B372} (1996) 265;\\
M.~Fukugita, M.~Tanimoto and T.~Yanagida, \PR{D57} (1998) 4429. 

%%% FMTY paper %%%
\bibitem{Fukuura_etal:2000}
K.~Fukuura, T.~Miura, E.~Takasugi and M.~Yoshimura,
\PR{D61} (2000) 073002.

%%% Tri-Maximal %%%
\bibitem{TriMax} N.~Cabibbo, \PL{72B} (1978) 333;\\
L.~Wolfenstein, \PR{D18} (1978) 958;\\
V.~Barger, K.~Whisnant and R.J.N.~Phillips, \PR{D24} (1981) 538;\\
A.~Acker, J.G.~Learned, S.~Pakvasa, and T.J.~Weiler, \PL{B298} (1993)
149;\\
R.N.~Mohapatra and S.~Nussinov, \PL{B346} (1995) 75;\\
P.F.~Harrison, D.H.~Perkins and W.G.~Scott, \PL{B349} (1995) 137; 
\PL{B374} (1996) 111; \PL{B396} (1997) 186; \PL{B458} (1999) 79; \\
C.~Giunti, C.W.~Kim and J.D.~Kim, \PL{B352} (1995) 357;\\
R.~Foot, R.R.~Volkas and O.~Yasuda, \PL{B433} (1998) 82.

%%% CP phases in Majorana neutrino %%%
\bibitem{Bilenky-etal1980} S.M.~Bilenky, J.~Ho$\check{\rm s}$ek and
  S.T.~Petcov, \PL{94B} (1980) 495;\\
%\bibitem{Doi-etal1981} 
M.~Doi, T.~Kotani, H.~Nishiura, K.~Okuda and
  E.~Takasugi, \PL{102B} (1981) 323.


\end{thebibliography}
\end{document}